\documentclass[iop]{emulateapj}

\usepackage[latin1]{inputenc}
\def\t9{T$_{90}$}

\def\nuf{\nu f}

\newcommand{\grb}{GRB 111209A~}

\usepackage{graphicx}

\shorttitle{The ultra-long GRB 111209A}
\shortauthors{Gendre et al.}

\begin{document}

%% LaTeX will automatically break titles if they run longer than
%% one line. However, you may use \\ to force a line break if
%% you desire.

\title{The ultra-long Gamma-Ray Burst 111209A: the collapse of a blue supergiant?}

%% Use \author, \affil, and the \and command to format
%% author and affiliation information.
%% Note that \email has replaced the old \authoremail command
%% from AASTeX v4.0. You can use \email to mark an email address
%% anywhere in the paper, not just in the front matter.
%% As in the title, use \\ to force line breaks.

\author{B. Gendre\altaffilmark{1}}
\affil{ASI Science Data Center, via Galileo Galilei, 00044 Frascati, Italy\\
Osservatorio Astronomico di Roma, OAR-INAF}
\email{bruce.gendre@gmail.com}

\author{G. Stratta}
\affil{Osservatorio Astronomico di Roma, OAR-INAF, via Frascati 33, 00040, Monte Porzio Catone, Italy}

\author{J.L. Atteia}
\affil{Université de Toulouse; UPS-OMP; IRAP; Toulouse, France\\
CNRS; IRAP; 14, avenue Edouard Belin, F-31400 Toulouse, France
}

\author{S. Basa}
\affil{Aix Marseille Université, CNRS, LAM (Laboratoire d'Astrophysique de Marseille) UMR 7326, 13388, Marseille, France}

\author{M. Bo\"er}
\affil{CNRS; ARTEMIS, UMR 7250; Boulevard de l'Observatoire, BP 4229, 06304 Nice Cedex 4, France\\
CNRS; Observatoire de Haute-Provence, 04870 Saint Michel l'Observatoire, France
}

\author{D. M. Coward}
\affil{University of Western Australia, School of Physics, University of Western Australia, Crawley WA 6009, Australia}

\author{S. Cutini}
\affil{1ASI Science Data Center, via Galileo Galilei, 00044 Frascati, Italy}

\author{V. D'Elia}
\affil{1ASI Science Data Center, via Galileo Galilei, 00044 Frascati, Italy}

\author{E. J Howell}
\affil{University of Western Australia, School of Physics, University of Western Australia, Crawley WA 6009, Australia}

\author{A. Klotz}
\affil{Université de Toulouse; UPS-OMP; IRAP; Toulouse, France\\
CNRS; IRAP; 14, avenue Edouard Belin, F-31400 Toulouse, France\\
CNRS; Observatoire de Haute-Provence, 04870 Saint Michel l'Observatoire, France
}

\and

\author{L. Piro}
\affil{Istituto di Astrofisica e Planetologia Spaziali di Roma, via fosso del cavaliere 100, 00133 Roma, Italy/INAF}

%% Notice that each of these authors has alternate affiliations, which
%% are identified by the \altaffilmark after each name.  Specify alternate
%% affiliation information with \altaffiltext, with one command per each
%% affiliation.

\altaffiltext{1}{On personal leave}

%% Mark off your abstract in the ``abstract'' environment. In the manuscript
%% style, abstract will output a Received/Accepted line after the
%% title and affiliation information. No date will appear since the author
%% does not have this information. The dates will be filled in by the
%% editorial office after submission.

\begin{abstract}
We present optical, X-ray and gamma-ray observations of GRB 111209A, observed at a redshift of z = 0.677. We show that this event was active in its prompt phase for about 25000 seconds, making it the longest burst ever observed. This rare event could have been detected up to z $\sim 1.4$ in gamma-rays. Compared to other long GRBs, GRB 111209A is a clear outlier in the energy-fluence and duration plane. The high-energy prompt emission shows no sign of a strong black body component, the signature of a tidal disruption event, or a supernova shock breakout. Given the extreme longevity of this event, and lack of any significant observed supernova signature, we propose that GRB 111209A resulted from the core-collapse of a low metallicity blue super giant star. This scenario is favoured because of the necessity to supply enough mass to the central engine over a duration of thousands of seconds. Hence, we suggest that GRB 111209A could have more in common with population III stellar explosions, rather than those associated with normal long gamma ray bursts.
\end{abstract}

%% Keywords should appear after the \end{abstract} command. The uncommented
%% example has been keyed in ApJ style. See the instructions to authors
%% for the journal to which you are submitting your paper to determine
%% what keyword punctuation is appropriate.


\section{Introduction}

Gamma-ray bursts (GRBs) are detected as brief flashes of high-energy photons typically lasting some tens of seconds \citep{kle73}. First discovered in the 1960s, these cosmological sources are the most energetic explosions in the Universe since the Big Bang \citep[see][for a review]{mes06}. There are two main classes of GRBs with well separated properties and behaviors: long and short GRBs \citep{kou93}.
Because of their extreme distances and diverse emission characteristics, it has been difficult to isolate a single progenitor responsible for all long duration GRBs. However, the association of rare type Ic supernovae with several long GRBs \citep[e.g.][]{hjo03} suggests that these GRBs indicate the formation of compact objects following stellar collapse. This hypothesis, based on the Collapsar model \citep{woo93}, suggests that the progenitor of long duration GRBs are Wolf-Rayet stars \citep[e.g.][]{che99}. In addition, other astrophysical phenomena can mimic GRBs, such as the magnetic reconnection onto a neutron star \citep{uso92} or the tidal disruption of a minor body by a compact object \citep{boe89}. These events must occur at closer proximities than classical long GRBs as their available energy is lower.

We here present an analysis of the longest burst ever observed, \grb, which had a duration of 25000 seconds. This burst occurred at a redshift of \emph{z} = 0.677 \citep{vre11}. Our analysis uses the observations of XMM-Newton \citep{jan01} and Swift \citep{geh04} in the X-ray, Konus-Wind \citep{apt95} in the gamma-ray band, TAROT \citep{klo09} in the optical, as well as archived and public data. In this paper we focus our attention on the prompt emission of GRB 111209A to determine the most likely astrophysical source of such an ultra-long event. An additional paper will discuss the afterglow properties in a multi wavelength context (Stratta et al., in preparation).

In the following, we define the classes of bursts with durations greater than 2\,s, $10^3$\,s, and $10^4$\,s as long, super-long and ultra-long GRBs respectively. We also define T$_{90}$ the time required to radiate 90\% of the total energy observed in a given band. We assume flat $\Lambda$ CDM Universe with $H_0 = 71$ km s$^{-1}$ Mpc$^{-1}$, $\Omega_m = 0.27$, and quote all errors at the 90 \% confidence level. The outline of the paper is as follows.

In Sections \ref{sec_obs} and \ref{sec_reduction} we present the observations and the data reduction respectively. The data analysis is described in Section \ref{sec_anal}. We discuss the duration and scarcity of events like \grb in Section \ref{sec_discu} and in Section \ref{sec_progenitor} consider plausible scenarios for the progenitor of \grb. We present our conclusions in the final Section 7.

\section{GRB 111209A}
\label{sec_obs}

\grb was discovered by the Swift satellite at $T_0$ = 2011:12:09-07:12:08 UT \citep{hov11}. This event initiated two triggers of the Burst Alert Telescope (BAT; triggers number 509336 and 509337). It showed an exceptionally long duration, was monitored up to $T_0+1400$ s. However, the burst started about 5400 seconds before $T_0$, as shown in the ground data analysis of the Konus-Wind instrument, and a re-analysis of the BAT data showed \grb was detectable  150 seconds before the trigger. Swift did not trigger at the start of the event as the burst was not in the field of view of the BAT instrument.

\grb~was also detected by Konus-Wind. In order to compare the light curve of this event with other bursts, we used the publicly available soft band (21-83 keV; similar to the Swift-BAT 15-150 keV band) light curve. The earliest portion of the gamma-ray signal detected by Konus-Wind featured a weak broad pulse. The gamma-ray signal was then observed as a multi-peaked emission up to about $T_0$ + 10000 s \citep{gol11}. Using the Konus-Wind results, GRB 111209A had a fluence of $(4.86 \pm 0.61) \times 10^{-4}$ erg cm$^{-2}$, an isotropic energy E$_{\rm iso} = (5.82 \pm 0.73) \times 10^{53}$ erg, and an intrinsic peak of the spectrum in the $\nuf_\nu$ space of E$_p = 520 \pm 89$ keV \citep{gol11}. These values are in agreement to within two sigma with the Amati relation \citep{ama02} which empirically links these last two quantities for long GRBs.

\begin{figure*}
\epsscale{1.0}
\plotone{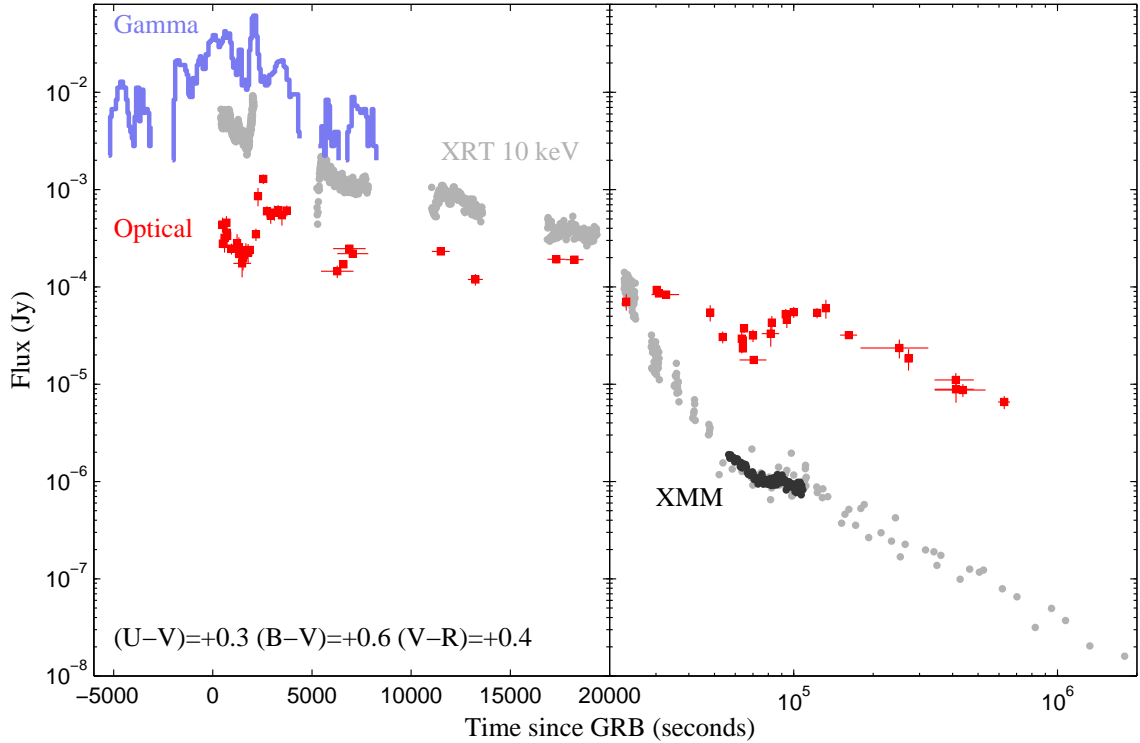}
\caption{The light curve of GRB 111209A, presented with a temporal axis which is linear in the left panel (the prompt emission) and logarithmic in the right panel (the afterglow emission). X-ray data are in grey (XRT) and black (XMM-Newton). The Konus-Wind data (blue solid line) has been scaled to the X-ray data for comparison. The R-band light curve is constructed from TAROT and Swift-UVOT data (color indices are indicated on the plot).\label{fig1}}
\end{figure*}

Swift/XRT observations started 425 s after the BAT trigger \citep{hov11} revealing a bright afterglow, observed also by Swift/UVOT in the optical-UV bands at RA(J2000)=00$^h$ 57$^m$ 22.63$^s$ and Dec(J2000)=-46d 48$'$ 03.8$''$, with an estimated uncertainty of 0.5 arcsec. The afterglow was also clearly detected by ground based instruments; for example the TAROT-La Silla \citep{klo11}, and the GROND robotic telescopes \citep{kan11}. In addition, we activated a Target of Opportunity observation with XMM-Newton, between $T_0 + 56,664$ s and $T_0 + 108,160$ s (see the light curve in Fig. \ref{fig1}). This period covered the end of the prompt phase seen in X-ray, a subsequent plateau phase, and the start of the normal afterglow decay \citep{gen11}.

%Finally, at $T_0$ + 1.9 days and $T_0$ + 5.1 days, the Australia Telescope Compact Array (ATCA) observed the position of \grb. The instrument failed to detect it during the first observation, with a 34 GHz 3$\sigma$ upper limit of 132$\mu$Jy \citep{han11}, but detected it on the second one at frequency of 5.5, 9, and 18 GHz \citep{han11b}.

\section{Data reduction}
\label{sec_reduction}

\subsection{Swift XRT}

Swift/XRT data were taken as Level 1 event files from the Swift Data Center archive at NASA/GSFC and were processed with the task {\em xrtpipeline} version 0.12.6, applying the most updated calibrations. Data acquisition started in Windowed Timing (WT) mode and continued in this mode up to the fourth orbit (T$_0$ + 22700 s). XRT then switched to the Photon Counting (PC) mode for the remaining of the observation (up to 26 days after the trigger). Pile-up can be observed in WT mode during the first orbit (i.e. up to T$_0$ + 2000 s), and in PC between $T_0$ + 22700 s and $T_0$ + 30900 s. As a consequence, spectra and light curves were extracted using the methods of \citet{rom06} and \citet{vau05}, excluding a central circle of radius 1 pixels from the PSF in WT mode; for the PC mode, the radius was set dynamically betwen 7 and 1 pixels as a function of the flux of the afterglow, and the results checked with the PSF profile of the XRT.

All spectra and light curves were then extracted using standard filtering and screening criteria (including bad columns and bad pixels), and corrected for telescope vignetting and aperture filtering using the ancillary response files generated by the task {\em xrtmkarf}. We then rebinned all files to either match the spectral resolution when the data quality was high enough, or to obtain at least 30 counts per bin.

\subsection{XMM-Newton}

We retrieved the Observation Data Files from the XMM archive and reprocessed them using the XMM-SAS version 12.0.1 and the latest available calibration files associated with this version of the SAS. The raw events files were processed using the tasks {\em emchain, epchain, rgsproc, omichain, omfchain}. The PN instrument observed GRB 111209A in Full Window mode, while the MOS cameras setting was on Small Window. Because of the brightness of the afterglow, we checked for pile-up in the data. We used the task {\em epatplot} for this purpose, and confirmed that the data were free from pile-up, even if a significant fraction of {\em out off time events} were observed. Because of this latter effect, the background regions have been chosen, when possible, far away from the source and corrected for the difference of off-axis angle using the ARFs. In the special case of the PN instrument, the instrumental background is highly variable depending on the position on the instrument -- this was taken into account by using a region where the instrumental background was compatible with the position of the source.

We then checked for flares of solar protons, that can induce a huge increase of the background count rate, and found some marginal events where the background rate was increased by a factor of 5. Due to the brightness of the afterglow, these events are not very significant for the following analysis: when maximal, the background represents around 6\% of the total signal at the position of the afterglow.

The event files were then filtered for good photons, using FLAG == 0 and Pattern filtering (less than 4 for the PN, less than 12 for the MOS). Spectra were extracted using the task {\em especget}, light curves with the task {\em evselect}.

\section{Data analysis}
\label{sec_anal}

\subsection{Temporal analysis}

\grb X-ray emission indicates an initial shallow decay phase. Assuming a power law model between $T_0$ + 0.425 ks and $T_0$ + 10.4 ks, the best fit decay index is $\alpha_p=0.544\pm0.003$. This phase ends very late in comparison with typical X-ray emissions. At the end of this phase, a gradual steepening drives the light curve to the so called ``steep decay" phase, thought to indicate the end of the prompt emission and is characterized by very fast decrease of the count rate. During this phase and the following one, the light curve can be modeled with a double broken power law with a steep decay index of $\alpha_s=4.9\pm0.2$, a plateau with decay index of $\alpha_f = 0.5\pm0.2$, and a final decay index of $\alpha_a=1.51\pm0.08$.

\subsection{Spectral analysis}

In the following, all models fitted to the data were multiplied by two components in order to take into account our Galaxy and the host galaxy photoelectric absorption from metals. The equivalent hydrogen column density of our Galaxy was set to $N_H = 1.48 \times 10^{20}$ cm$^{-2}$ \citep{kal05}. The second absorption component was left free to vary at the redshift of the burst, assuming solar metalicity. We found no temporal variation of the intrinsic $N_H$, with a mean of $(2.5 \pm 0.4) \times 10^{21}$cm$^{-2}$.% and in the following fix it to $2.5 \times 10^{21}$cm$^{-2}$ to reduce the uncertainties on the other measured parameters.

We began with an analysis of the prompt spectral emission between $T_0$ and $T_0 + 2500$ s, seen by Swift/XRT. \grb shows a strong flare at $T_0 + 1200$ s. Since flares are known to have highly variable spectra, this was excluded in order to investigate the underlying spectral continuum. The spectrum was of exceptionally high quality, and was re-binned in order to obtain at least 400 counts per bin. The background subtracted energy spectrum between $T_0 + 425$ s and $T_0 + 1000$ s is inconsistent with a simple power law or cut-off power law model. The best fitting model was that based on a broken power-law added to a blackbody component. The high energy segment of the broken power-law is consistent with the BAT spectrum at the same time \citep[$\Gamma = 1.48 \pm 0.03$;][]{pal11}, but harder than the time averaged Konus-Wind best fit model. We note however that the following spectra (extracted between 5.0 and 47.8 ks) are softer, consistent with the time averaged Konus-Wind spectrum. We therefore assume that the spectral position of the break (i.e. $E_{break}$) varies with time, and that this break is due to the crossing of a specific frequency within the observed band. The black body component is very soft (see Table \ref{table_fit}), and accounts for about 0.01 \% of the total flux in the 0.5-10.0 keV band. Its luminosity is $\sim 1.6 \times 10^{50}$ erg s$^{-1}$. Moreover, this component is not detected at later times.

\begin{deluxetable*}{cccccccc}
%\rotate
\tablecaption{X-ray spectral analysis of GRB 111209A.\label{table_fit}}
\tablehead{
\colhead{Instrument} & \colhead{Start} & \colhead{End} & \colhead{Spectral} & \colhead{$\Gamma$} & \colhead{$E_{0}$} & \colhead{Second Spectral} &  \colhead{$\chi_{\nu}^2$}\\
\colhead{observing} & \colhead{time} & \colhead{time} & \colhead{Model} & \colhead{} & \colhead{} & \colhead{component} & \colhead{(dof)}\\
\colhead{mode} & \colhead{(s)} & \colhead{(s)} & \colhead{} & \colhead{} & \colhead{(keV)} & \colhead{(eV or energy index)} &  \colhead{}
}
\startdata
Swift/WT    & 425   & 1000   & BKN+BB & $1.17\pm0.03$       & $3.4 \pm 0.4$   & $30 \pm 5$          &  0.98 (181) \\
            &       &        &        & $1.52\pm0.06$       &                 &                      &             \\
Swift/WT    & 5025  & 7825   & PL     & $1.58\pm0.03$       &                 &                      &  1.13 (590) \\
Swift/PC    & 22700 & 25130  & PL     & $1.8\pm0.1$         &                 &                      &  0.87 (50)  \\
Swift/PC    & 28780 & 30880  & PL     & $1.9\pm0.1$         &                 &                      &  0.78 (32)  \\
XMM-Newton	&	55073 & 61416	 & 2PL    & $0.9 \pm 0.7$       &                 & $2.7\pm0.2$          &  0.89 (151) \\
XMM-Newton	&	61416 & 66416	 & 2PL    & $1.2^{+0.7}_{-0.9}$ &                 & $2.8^{+0.4}_{-0.2}$  &  0.84 (111) \\
XMM-Newton	&	66416 & 71416	 & 2PL    & $1.9^{+0.4}_{-0.8}$ &                 & $3.3^{+0.9}_{-0.5}$  &  0.86 (95)  \\
XMM-Newton	&	71416 & 76416	 & 2PL    & $1.7^{+0.5}_{-0.9}$ &                 & $3.1^{+0.8}_{-0.4}$  &  1.05 (83)  \\
XMM-Newton	&	76416 & 91416	 & 2PL    & $1.7^{+0.5}_{-0.7}$ &                 & $3.0^{+0.5}_{-0.3}$  &  0.89 (228) \\
XMM-Newton	&	91416 & 108599 & 2PL    & $2.3^{+0.3}_{-0.8}$ &                 & $3.4^{+1.3}_{-0.6}$  &  1.04 (234) \\
\enddata
\end{deluxetable*}

%
%_____________________________________________________________

%______________________________________________ Gamma_1 (lg rho, lg e)
   \begin{figure}
   \centering
   \epsscale{1.0}
   \plotone{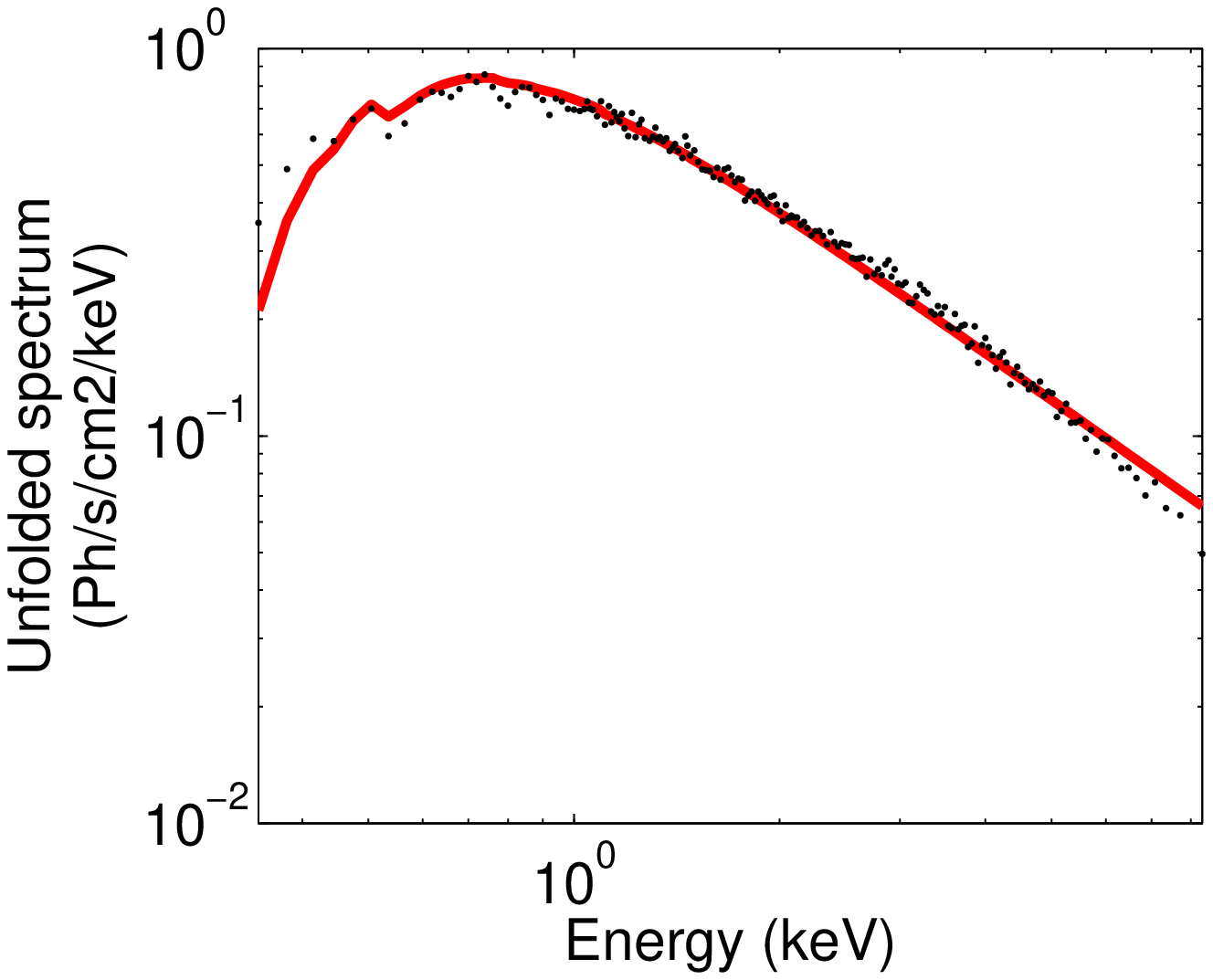}
   \plotone{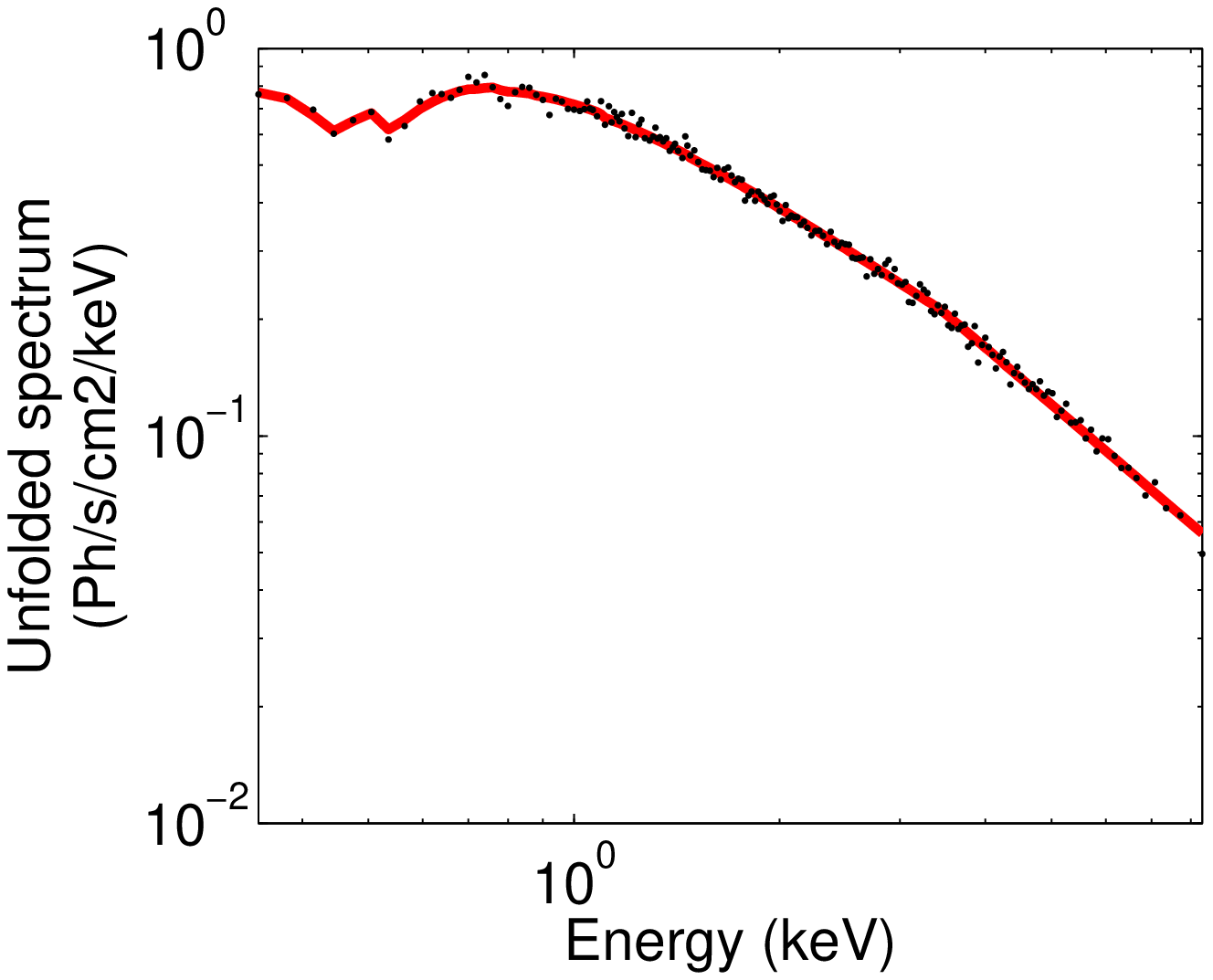}
   \caption{Swift/XRT WT spectrum. The spectrum is fitted with a simple power law (upper panel) or with the addition of a black body and a broken power law (lower panel). An improvement fit is clearly visible in the bottom panel.\label{figure:1orb}}
    \end{figure}

The XMM-Newton monitoring period covered the transition from the steep decay phase to the ``normal" decay phase, for an interval of 53 000 seconds. Because of the greater sensitivity of XMM-Newton in comparison with the Swift XRT, we were able to extract the spectra in six different time intervals (see Tab. \ref{table_fit}). A simple power law model was first used, giving an acceptable value of the $\chi^2$ statistic value. However, the residuals of the fit clearly showed systematic excesses at the hard end of the spectrum, and thus a double component model was preferred. The inclusion of a second component produced an F-test probability (the probability that the decrease of the $\chi^2$ statistic value is due to a spurious effect) of the order of $10^{-9}$. The spectral fit clearly indicates that this second component is not dominating, and provides a small contribution to the total observed flux.

We extracted a Spectral Energy Distribution (SED) during the XMM-Newton observation using published results from \citet{kan11} in the optical, in order to confirm no thermal emission was present during this phase. \citet{kan11} already noted that the GROND data excluded a thermal component, and indeed we have found an optical spectral index of $\beta = 1.1 \pm 0.2$, using the optical data alone. Using the X-ray to optical data, we found that an absorbed broken power law fitted the data well. We note that we can exclude the presence of a thermal emission in the far UV or the soft X-ray parts of the broad band spectrum.

As stated previously, this paper focuses on the properties of the progenitor of \grb, and the nature of this second component will be investigated in more detail in a forthcoming paper (Stratta et al., in preparation). We note however that this is the first time, to date, that such a double non-thermal component has been detected during the steep decay phase. Several models can account for it \citep[e.g. see the model of][]{bar09}. The optical light curve features a rebrightening during the XMM-Newton observation: this component may be linked to the prompt-to-afterglow transition, to the optical bump, or even be totally unrelated to these components.

\section{Discussion}
\label{sec_discu}

\subsection{Duration of the prompt emission}

The estimation of the burst duration using classical methods (i.e. T90) is complicated by two factors: it is well known that the duration depends on the observing band (the softer the band, the longer the detected signal); and for this event the prompt phase is split into several orbits, with gaps in the signal. Significantly, the event was recorded by Konus-Wind before $T_0$, when the burst was not in the field of view of Swift/BAT. Therefore, all attempts to use classical methods are flawed. To give an estimate of the total duration for comparison with other bursts, we thus estimated the burst start and stop epochs $T_{start}$ and $T_{end}$  as follows:
\begin{itemize}
\item For $T_{start}$, we use the start of the detection by Konus-Wind, about 5400 seconds before $T_0$;
\item For $T_{end}$, we use the time of the start of the sharp decay, about 20,000 seconds after $T_0$.
\end{itemize}

The choice of the $T_{end}$ value is motivated by the fact that the sharp decline is usually interpreted as the high latitude emission of the prompt phase, and its start time corresponds to the true end of the prompt phase \citep{kum00}. Before the steep decay phase, the initial XRT and BAT light curves are usually well correlated, indicating a common origin. This is also the case for \grb (see Fig. \ref{fig1}). With these assumptions, the total duration of \grb is at least 25,000 seconds.

We extensively surveyed the Swift, BATSE and Konus-wind catalogs to find other ultra-long bursts. The most complete and homogeneous catalog of prompt burst properties is the BATSE 4B catalog \citep{pac99}. From this catalog, only few bursts have a duration larger than 1000 seconds (the super-long ones), and none have a duration longer than 10 000 seconds. For the case of Swift, we also searched in the XRT light curve repository \citep{eva07} to obtain the start time of the sharp decay for all bursts, and found no burst with such a large duration. Thus, GRB 111209A is to date the longest confirmed GRB ever observed, and its duration is exceptional.

We note however, for completeness, that one event, longer than GRB 111209A, has a debated origin. Swift \object{J164449.3+573451} (hereafter J1644+57) has been proposed by several groups to be a tidal disruption event \citep[e.g.][]{bur11, blo11, lev11, tch13}. Nevertheless \citet{qua12}, even if agreeing that the most plausible origin was a tidal disruption event, have shown theoretically that the explosion of a massive star could produce a similar event.

\subsection{Selection effects on ultra-long GRBs}

It is possible that selection effects make only the brightest part of ultra-long burst light curves detectable, thus reducing their intrinsic durations and making ultra-long GRBs indistinguishable from normal bursts. We investigated this selection effect in two ways.

\begin{figure}
\epsscale{.80}
\plotone{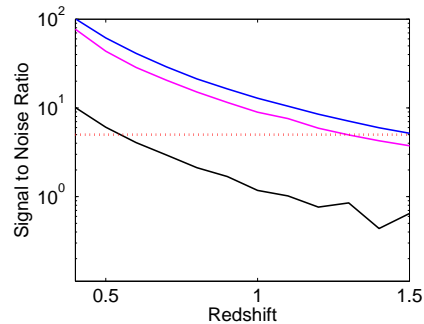}
\caption{Detection significance expressed in signal to noise ratio units for a GRB 111209A like event as a function of its redshift. We present the detection regime of such bursts using light curve bin sizes of 1 minute (purple line) and 2 minutes (blue line); and a burst 10 times fainter with 2 minutes bin size (black line). The dotted red line indicates our detection threshold (5 sigma level), showing that even in the optimal case the maximum redshift for detection is \emph{z} =  1.4-1.5.\label{figS3}}
\end{figure}

Firstly, we constrained the maximum distance from which this burst could be detected. We extracted data from the Konus-Wind light curve, re-scaling it to a given distance and then added the background of the instrument (adding statistical fluctuations in the final signal to take into account Poisson statistics). We used two signal levels: the original one, and a second level an order of magnitude fainter. The light curve was examined for a burst with a rate trigger threshold of five sigma. We used two different bin sizes for the light curve (1 and 2 minutes), the latter simulating more complex trigger methods where an image trigger can be considered. The simulation was performed 100 times for all redshifts between 0.3 and 2.2 (with a step of 0.1). Results are presented in Fig. \ref{figS3}, and are quite clear: GRB 111209A cannot be detected at high-z. In fact, this event can be detected only within a limiting redshift of z = 1.4.

\begin{figure}
\epsscale{.80}
\plotone{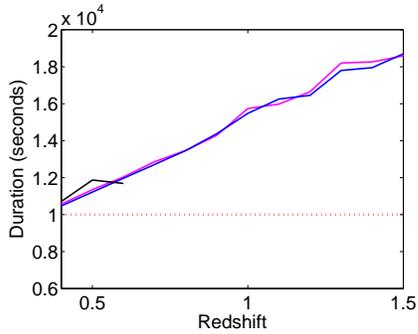}
\caption{Measured duration of a GRB 111209A like event as a function of its redshift. Colors are the same as those in Fig. \ref{figS3}. The horizontal red dotted line represent 10 000 seconds. The black line is truncated at the point where the errors are larger than the measured duration.\label{figS4}}
\end{figure}

Secondly, for positive detections (at $z < 1.4$), we computed the expected duration of the burst, using the standard GRB procedure to obtain T90. Again, this is done for two distinct temporal resolutions and flux thresholds. Results are presented in Fig. \ref{figS4}. Surprisingly, we find that the duration is larger than 10000 seconds in all cases. Thus, an ultra-long burst that could trigger an instrument would always be detected as an ultra-long GRB. The lack of such other bursts in $\sim$ 30 years of archives indicates a very rare event in the local Universe. Such events may have been more common in the distant Universe, but their faintness makes their detection unfeasible.

\subsection{The rate of ultra-long GRB in the local Universe}

We estimated the rate of ultra-long GRBs in the local Universe using the properties of \grb. Taking into account the effect of the redshift on Swift's sensitivity and k-correction \citep{cow12,how13}, this rate density can be inferred from the measured peak flux. We used the brightest epoch of the GRB 111209A prompt phase, as seen by Konus-Wind, scaled to the Swift energy band. This is then scaled to the limiting detection flux of Swift to obtain a maximal redshift for detection for Swift, $z_{lim}$ \citep[see][for details]{cow12}. The corresponding volume of detectability is then :

\begin{equation}
V_{max} = \int^{z_{lim}}_{0} \frac{dV}{dz}dz
\end{equation}

Using the cosmological model provided in the Introduction, the maximum detectable volume corresponding to $z_{lim}=1.4$ is $V_{\rm{max}} = 292$ Gpc$^{3}$. Following \citet{gue07,cow12} we calculate a rate density:

\begin{equation}\label{eq_vmax_slgrb}
    R =  \frac{1}{V_{\rm{max}}} \frac{1}{T} \frac{1}{\Omega} \frac{1}{\eta_{z}}\,,
\end{equation}

\noindent with $T$ the maximum observation-time for the sample (7 years), $\Omega=0.17$ the sky coverage of the instrument and  $\eta_{z}$ =0.3 the fraction of \emph{Swift} bursts with a measured redshift. We obtain $9^{+ 27}_{-8} \times 10^{-3}$ Gpc$^{-3}$ yr$^{-1}$ where the errors are the 90\% Poisson confidence limits \citep{geh86}.

As only four out of 644 Swift bursts detected between the launch of the satellite and March 2012 have prompt emission durations greater than some thousands of seconds, we obtain the final rate for ultra long GRBs to be of the order of $4 \times R / 644 \sim 6 \times 10^{-5}$ Gpc$^{-3}$ yr$^{-1}$. This is significantly lower than that of the normal long GRB rate density \citep[see][]{how13}.

\subsection{Comparison with super-long bursts}

\begin{figure}
\epsscale{1.2}
\plotone{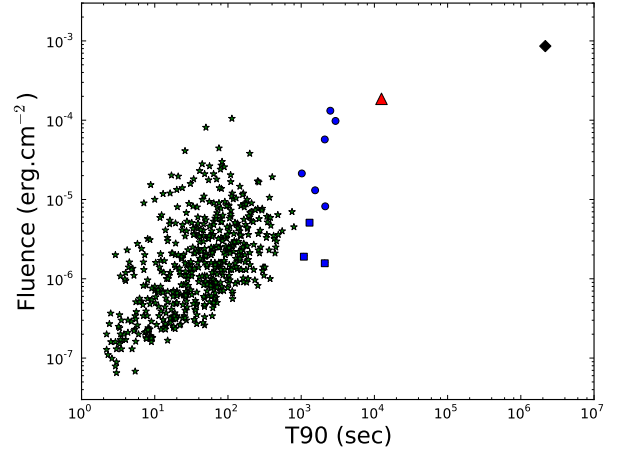}
\caption{Position of GRB 111209A (red triangle) in the GRB Fluence-Duration plane. The green stars are long Swift bursts. Blue circles and squares are the recorded super long GRBs, and the black dot is J1644+57. Blue squares represent high-energy transients attributed to SN shock breakout. Note that all measurements are converted to the Swift band (15-150 keV) for comparison with the Swift catalog - therefore some values will differ from those given in Table \ref{table_s1}. \label{fig3}}
\end{figure}

\begin{figure}
\epsscale{1.20}
\plotone{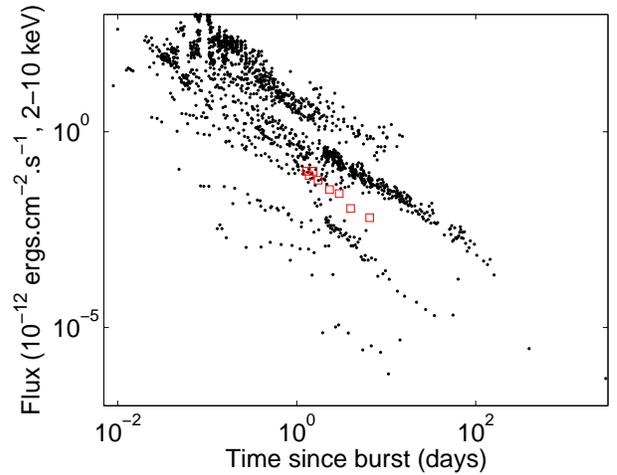}
\caption{The absolute flux of GRB 111209A. Following the method of \citet{gen08a}, we rescale the X-ray afterglow light curve of GRB 111209A to a common distance of z = 1. This light curve is represented by red squares, while the sample of \citet{gen08a} are shown as black dots.\label{figS2}}
\end{figure}

We retrieved from the literature the prompt emission information available for the 9 super-long bursts (with duration larger than 1000 seconds) listed in Table \ref{table_s1}. The fact that GRB 111209A is harder and more fluent than other bursts is another indicator of its uniqueness. In Fig. \ref{fig3} we also present these bursts in the duration-fluence plane. \grb appears as an outlier in the duration distribution, but not in the fluence distribution, where other super-long GRBs can have a similar (even if slightly smaller) fluence. We note that supernova shock breakouts have a much smaller fluence, weakening a possible link between this kind of progenitor and \grb. Once corrected for distance (when possible - several bursts lacked a good follow-up campaign and thus a redshift measurement), this burst also appears to be more energetic than the others.

\begin{deluxetable*}{ccccccc}
\tabletypesize{\scriptsize}
%\rotate
\tablecaption{Summary of the prompt properties of all bursts with duration larger than 1000 seconds. \label{table_s1}}
\tablewidth{0pt}
\tablehead{
\colhead{Quantity} & \colhead{E$_p$} & \colhead{E$_{\rm iso}$}   & \colhead{Fluence}         & \colhead{Duration} & \colhead{Distance} & \colhead{Reference}\\	
\colhead{        } & \colhead{(keV)} & \colhead{(10$^{51}$ erg)} & \colhead{(erg cm$^{-2}$)} & \colhead{(T90, s)} & \colhead{(redshift)} & \colhead{}
}
\startdata
GRB 971208  & 144    & n.a.    & $2.6 \times 10^{-4}$ & 2500     & n.a.   & \citet{pal08} \\
GRB 020410  & n.a.   & n.a.    & $2.8 \times 10^{-5}$ & 1550     & n.a.   & \citep{nic04} \\
GRB 060218  & 4.9    & 0.062	 & $1.7 \times 10^{-5}$	& 2100	   & 0.0331 & \citet{cam06} \\
GRB 060814B & 341    & n.a.	   & $2.4 \times 10^{-4}$ & 2944     & n.a.	  & \citet{pal08} \\
GRB 080407  & 287    & n.a.	   & $4.5 \times 10^{-4}$	& 2100	   & n.a.	  & \citet{pal12} \\
GRB 090417B & $>150$ & $> 7.3$ & $8.2 \times 10^{-6}$	& $> 2130$ & 0.345	& \citet{hol10} \\
GRB 091024  & 280    & 350	   & $1.1 \times 10^{-4}$	& 1200	   & 1.09	  & \citet{gol09} \\
GRB 100316D & 10-42  & 0.049	 & $5.1 \times 10^{-6}$	& 1300	   & 0.0591	& \citet{sta11} \\
GRB 101225A & 38     & n.a.	   & $2.6 \times 10^{-6}$ &	$> 2000$ & n.a.	  & \citet{cam11, tho11} \\
Swift J1644+57 & n.a.   & n.a.    & $8.6 \times 10^{-4}$ & 2160000  & 0.354  & \citep{bur11} \\
GRB 111209A & 520    & 580     & $4.9 \times 10^{-4}$ & $>25000$ & 0.677	& This work \\
\enddata
\end{deluxetable*}

We also investigated the X-ray afterglow properties of these bursts. As shown in previous works using infrared, optical and X-ray data, the luminosity of afterglows tends to cluster in several groups \citep{boe00, gen05, lia06, nar06, gen08a, gen08b}. If one includes the X-ray light curve of GRB 111209A in the sample of \citet[][see fig. \ref{figS2}]{gen08a}, the event is shown to be marginally under luminous. However, this could be interpreted as a normal consequence of the duration of the prompt phase. In this phase, internal shocks dissipate the energy of the fireball. The longer the prompt phase lasts, the less energy remains to be dissipated within the following external shock.

\section{Nature of this event}
\label{sec_progenitor}

In the following section, we investigate the possible progenitors of such an ultra-long event.

\subsection{Thermal components}

The first hypothesis is that the exceptional duration of \grb~is due to a component usually not present in normal GRBs. Two super long GRBs (GRB 060218, z = 0.033, and GRB 100316D, z = 0.059) were associated with a supernova shock breakout \citep{cam06, sta11}. These two events are also classified as sub-energetic \citep{how13}, with very soft X-ray spectra, and their prompt spectra (including the steep decay) shows a thermal emission, which is responsible for the long duration. We also do observe a thermal component at the start of the XRT observation; however this component disappears very soon, and most of the prompt phase is free of thermal emission. We can therefore discard this hypothesis.

\citet{sta12} have found in several GRBs associated with a supernova a thermal component. They explain it as a hot plasma cocoon associated with the jet. However, they observe this component in the steep decay phase (where we do not see any thermal emission from GRB 111209A using XMM-Newton data) which is harder and dimmer than the one we detect at the start of the prompt phase. Thus, this model cannot be applied to GRB 111209A.

GRB 101225A, another super-long burst \citep{rac10}, shows a strong optical thermal component. The lack of a redshift measurement for this burst made difficult to draw any conclusions on the explosion mechanism and two very different models were proposed: a minor-body tidal disruption by a neutron star in our galaxy \citep{cam11}, and the merger of a neutron star with a helium star in a distant galaxy \citep{tho11}.

Apart from the lack of any strong thermal emission at optical (and X-ray) wavelengths, GRB 111209A differs from GRB 101225A as it is at least $10^3$ times more energetic \citep[using the most distant scenario of][]{tho11}, suggesting a different origin for GRB 111209A.

\subsection{Magnetars}

In the recent years, evidence has emerged that magnetars could account for long GRB properties \citep[e.g.][]{met10}. This class of progenitor provides a natural explanation of extended emissions as long lasting energy injection \citep{dal10, met11}. Indeed, if the typical duration of a long GRB powered by a magnetar is of the order of 40-100 seconds, atypical values of the magnetic field and spin period could allow a duration of $\sim 25,000$ seconds. However, such atypical values  cannot fit both the observed values of $E_{iso}$ and $E_p$. For instance, using the formalism of \citet{met11}, reproducing the observed values of E$_{iso}$ and duration would lead to E$_p \sim 180$ keV, that is a factor of 3 lower than the rest frame peak energy of GRB 111209A (520 keV). Trying to conciliate these values results in non physical solutions \citep[e.g. a neutron star spinning at a value exceeding the breakup value, $P_{lim} = 0.96$ ms,][]{lat04}.

Another possibility would be to consider the long duration of the prompt phase as a plateau phase (i.e. assume the true prompt phase has been missed by all instruments). However, the magnetar model indicates that once the plateau phase is finished the decay should be monotonic with a decay index of $1-2$ \citep{dal10}. This is in contradiction with our findings from the light curve (decay of $\alpha = 4.9 \pm 0.2$ with the presence of a following plateau).

\citet{tro07} has proposed a solution \citep[later extended by][]{lyo09} to this problem, with a two-stage progenitor: the collapsar would produce a magnetar that would collapse into a black hole a few hundreds of seconds later. Under this scenario one could expect a double plateau, i.e. a non-monotonic decay after the first plateau phase. The first plateau, due to the energy injection by the magnetar, would then be the observed prompt signal. However, again, one needs atypical values of the magnetic field and spin period to explain the observed properties of this event, and again this would produce inconsistent values of $E_{iso}$ and $E_p$, not observed in this event. Significantly, the ``first plateau" has also been observed in the gamma-ray band (it is the prompt signal), which is itself unique, making \grb~a very peculiar event.

We thus conclude that a magnetar progenitor for \grb is not supported by the data.

\subsection{Tidal Disruption Event}

Swift J1644+57 was at first claimed to be a GRB. It is thus possible that \grb is not a GRB but a tidal disruption event. This solution is however not very convincing. Firstly, no host galaxy has been detected, while one may expect a massive black hole to lie in a galaxy bright enough to be detected at small redshift. Secondly, the light curve should decay as $t^{-5/3}$ \citep{lod09}. This is however not the case (see Sec. \ref{sec_anal}). In fact, Swift J1644+57 did not show the characteristic GRB light curve behavior \citep[so called steep-flat-steep,][]{zha04} clearly visible in the case of \grb. For these reasons we can discard this hypothesis.

\subsection{Giant stars}

The main difficulty in explaining the nature of the progenitor of \grb~is its duration. \citet{woo12} investigated scenarios to provide a long duration event with an energy budget typical of a long GRB. They proposed four scenarios:
\begin{itemize}
\item i) single super-giant stars with less than 10\% solar metallicity;
\item ii) super-giant systems in tidally locked binaries;
\item iii) pair-instability in very high mass stars resulting in collapse to a black hole;
\item iv) helium stars in tidally locked binaries.
\end{itemize}

The scenarios ii) and iv) lead, assuming a typical values of 1\% $\dot{M}c^2$ for the energy conversion efficiency, to luminosities orders of magnitude too faint in comparison to the that observed in the case of \grb. Only a 100\% energy conversion efficiency would marginally reconciliate these numbers. As for the scenario iii), at a distance of \emph{z} = 0.677, one would expect a bright optical supernova easily detectable with large aperture telescopes. Despite follow-up of GRB 111209A by the Hubble Space Telescope and the Very Large Telescope, no evidence of a supernova was claimed \citep{lev12}. The above arguments leave us with scenario i), a single supergiant star with low metallicity. We note \citet{qua12} used the same kind of progenitor to discuss the link between Swift J1644+57 and a GRB.

The hypothesized progenitors of long duration GRBs are Wolf-Rayet stars (stars with the outer layers expelled during stellar evolution). When these layers are still present, as in low metallicity super-giant stars with weak stellar winds, the stellar envelope may fall-back and accretion can fuel the central engine for a much longer time. In this scenario, blue super-giant stars can produce GRBs with prompt emission lasting about $10^4$ seconds \citep{woo12}.

Theoretically, a larger star, such as a red super-giant star could also be considered, assuming that the outer layers are expelled in order to maintain the free-fall time scale to the order of $10^4$ seconds. For instance, a very weak explosion could eject the convective envelope of a red supergiant without producing an observable supernova, \citep[e.g.,][]{dex12}. However, in such a case, this layer of matter is still surrounding the progenitor of the GRB, making the surrounding medium density profile much more complex. As shown by e.g. \citet{che04}, these complex geometries should imprint their mark on the afterglow light curve, and in fact do \citep[see for instance GRB 050904,][]{gen05b}. Here we do not see this complex afterglow behavior, supporting the suggestion of a simple blue super-giant star.

If one considers the ratio of normal GRBs to ultra-long GRBs, this is shown in the Swift archives to be of order 200:1. In comparison, the fraction of blue super-giants compared to Wolf-Rayet stars is typically 10:1 \citep{eld08} -- therefore ultra-long GRBs should be more frequent than normal long GRBs. At first consideration, these numbers are difficult to reconcile. However, the greatest uncertainty of super-giant collapsars is whether matter with high angular momentum will be ejected by mass loss or will fall into the center of the star \citep{woo12}. It is plausible that for most of blue supergiant stars there is no fall back of the external layers, and thus no long emission. Hence, the observed and the expected rates may be reconciled by assuming the external layers of most blue super-giants do not fall into the stellar core. A rare combination of exceptionally large rotation and small mass loss at the end of stellar evolution could lead to a GRB 111209A like event. These properties can be both achieved in case of a low metallicity progenitor. This would also help reconcile the numbers, as most of the local universe blue super-giant stars are not observed with low metallicity. One should then consider if it is possible to have these low metallicity stars in the local Universe.

\subsection{Low metal protogalaxies in the local Universe}

The most putative hypothesis on the above consideration is the presence in the local Universe of a galaxy with a very low metallicity. In fact, blue supergiants have short lifetimes, and the only way of having a low metallicity blue supergiant is to form it from low-metallicity gas. We must thus discuss the possibility of having very low-metallicity galaxies in the local universe. If such galaxies should have been common in the early Universe, the hierarchical galaxy formation model implies that these have merged to form the larger galaxies observed today \citep[see e.g.][]{kly99}. However, as pointed out by several authors, such a mechanism is not efficient at 100 \% \citep[e.g.][]{kly99, die07}, and galaxies with a very low metallicity should still exist in the local Universe. In fact, several of those are known \citep{mor11}, and one object (I Zw 18) has a measured oxygen abundance of 0.05 solar one \citep{izo97}. These are local objects (with $z < 0.1$), accounting for about 0.1 \% of all the local galaxies \citep{mor11}, thus reinforcing the idea that \grb is a very rare event.

Furthermore, as noted by \citet{mor11}, the luminosity-metallicity relationship \citep{leq79} implies that low-metallicity galaxies cannot be observed in the more distant Universe. The fact that the host galaxy of GRB 111209A has not been resolved by the Hubble Space Telescope \citep{lev12} would then support this hypothesis: only the GRB afterglow was visible, and GRB 111209A traces the location of a putative (very) metal poor galaxy at large distance (z=0.677). At this distance, this galaxy would not have been detected without the GRB which occurred in it.

%the location of \grb would then be the location of the putative farthest extremely metal poor galaxy known so far.

\section{Conclusion}
\label{sec_conclu}

We have presented multi-wavelength observations of \grb. From the prompt emission, we have shown that this event is the longest GRB ever detected. We have also demonstrated that such a burst would have been detected up to $z=1.4$ as an ultra-long event. The lack of any ultra-long GRBs suggests that such events are very rare.

The spectrum shows a very soft thermal emission during the first thousands of seconds of the X-ray observations, not observed at later times. In addition, we have detected the presence of a puzzling second component seen in X-rays during the fast decay and the following plateau phase. This spectral signature is very uncommon and suggests an unusual nature for the progenitor of this burst.

Explaining these spectro-temporal properties is difficult in the standard collapsar model, and we propose that this event is representative of a new class of rare stellar explosion, that of a blue super giant star with a very low metallicity. We stress that the presence of such a low-metallicity giant star at this distance of our Galaxy, although rare, are not impossible. Thus the progenitor of this event could represent the closest link ever discovered with rare population III stellar explosions \citep{suw11, nak12}. One should also note that this event would not have been detectable at large redshift with the current instrumentation available. Only new instruments with very large detection areas, such as {\em LOFT} \citep{fer12} would be capable of such observations. However, at redshift 12, GRB 111209A would have a duration of $10^5$ seconds (i.e. more than one day) making such events very difficult to detect.

The data of \grb~are very rich, and this paper has only investigated the most prominent part of it. In a second paper, we will present a detailed study of the afterglow and the broad band spectrum, thus deducing the properties of the emission mechanism and of the surrounding medium.

\acknowledgments

We would like to thank the anonymous referee for his/her valuable comments, and E. Troja for discussions. This work has been financially supported by ASI grant I/009/10/0 and by the PNHE. D.M. Coward is supported by an Australian Research Council Future Fellowship. TAROT has been built with the support of the CNRS-INSU. We thank the technical support of the XMM-Newton staff, in particular N. Loiseau and N. Schartel.

{\it Facilities:} \facility{Swift}, \facility{XMM-Newton}, \facility{TAROT-Calern}, \facility{TAROT-La Silla}.

\clearpage

\end{document}